# Dissipative coupling induced phonon lasing with anti-parity-time symmetry

Qiankun Zhang[1], Cheng Yang[1], Jiteng Sheng[1,2,*], and Haibin Wu[1,2,3*]

[1]State Key Laboratory of Precision Spectroscopy, East China Normal University, Shanghai 200062, China

[2]Collaborative Innovation Center of Extreme Optics, Shanxi University, Taiyuan 030006, China

[3]Shanghai Research Center for Quantum Sciences, Shanghai 201315, China

*jtsheng@lps.ecnu.edu.cn; hbwu@phy.ecnu.edu.cn

**Phonon lasers, as the counterpart of photonic lasers, have been intensively studied in a large variety of systems, however, (all) most of them are based on the directly coherent pumping. Intuitively, dissipation is an unfavorable factor for gain in a laser. Here we demonstrate a novel mechanism of phonon lasing from the dissipative coupling in a multimode optomechanical system. By precisely engineering the dissipations of two membranes and tuning the intensity modulation of the cavity light, the two-membrane-in-the-middle system shows anti-parity-time (anti-PT) symmetry and the cavity mediated interaction between two nanomechanical resonators becomes purely dissipative. The level attraction and damping repulsion are clearly exhibited as the signature of dissipative coupling. After the exceptional point, a non-Hermitian phase transition, where eigenvalues and the corresponding eigenmodes coalesce, two phonon modes are simultaneously excited into the self-sustained oscillation regime by increasing the interaction strength over a critical value (threshold). In distinct contrast to conventional phonon lasers, the measurement of the second-order phonon correlation reveals the oscillatory and biexponential phases in the nonlasing regime as well as the coherence phase in the lasing regime. Our study provides a new method to study phonon lasers in a non-Hermitian open system and could be applied to a wide range of disciplines, including optics, acoustics, and quantum many-body physics.**

**One sentence summary: The dissipative coupling generates phonon lasing with anti-parity-time symmetry.**

Considerable interest in phonon lasers has arisen in recent years due to their wide range of potential applications in ultrasensitive sensing and information processing [1]. Especially, tremendous progress has been made in the self-sustained oscillation associated with radiation pressure in cavity optomechanics [2]. Most of phonon lasers demonstrated in a large variety of systems are based on a single lasing mode [3-9]. The recently realized multimode phonon lasers provide an opportunity of investigating many intriguing phenomena with multiple lasing modes [10-16], while most of them are used to study synchronization or phase-locking. To our knowledge, (all) most of the phonon lasers demonstrated, either single mode or multimode, stem from the coherent coupling, in which a directly coherent pumping is exploited with population inversion or nonlinear processes (e.g. Raman and Brillouin scattering), similar to an optical lasing process [6]. Other methods such as feedback and phase modulation have also been used to obtain coherence phonon signals [17,18]. In contrast to directly coherent coupling, dissipative coupling, which is owing to the indirect coupling to a common reservoir, could play a very important role in the open complex systems. Associated with non-Hermiticity and symmetries, it could lead to many extraordinary physics. Chiral dynamics [19] and energy-difference conservation [20] have been demonstrated in dissipatively coupled systems, however, dissipative coupling in a phonon laser has remained elusive.

In this work, we demonstrate a multimode phonon laser owing to the purely dissipative coupling in contrast to the directly coherent pumping. By precisely engineering a two-membrane-in-the-middle optomechanical system [21], the phonon lasing can arise from the purely dissipative coupling between two nanomechanical resonators, which is realized by allowing two mechanical resonators interacting indirectly through a common dissipative channel, i.e. the cavity field, via dynamical backaction. Such a system exhibits the non-Hermitian properties and could be exploited to investigate the phenomenon of anti-parity-time (anti-PT) symmetry breaking [20-25]. By increasing the dissipative coupling strength, the imaginary parts of the eigenvalues of the coupled system bifurcate after the system across the exceptional point (EP). Both resonators simultaneously enter into the phonon lasing regime once one of the hybridized modes exceeds the lasing threshold. The EP and phonon lasing threshold are clearly observed and tunable with the system parameters. The signature of strongly dissipative coupled systems is revealed as the level attraction and damping repulsion [26-28], which is distinct from the level repulsion for the coherent coupling. The second-order phonon correlation functions are measured to distinguish the different phases, which exhibit the oscillatory behavior before the EP, the biexponential phase between the EP and phonon lasing threshold, and the coherent properties after the lasing threshold. We also demonstrate that precisely tuning decay rates of resonators is critical to observe level attraction and anti-PT symmetry. When the effective decay rates of resonators are significantly different, a phenomenon analogous to electromagnetically induced absorption (EIA) is observed instead [29]. Our results reveal that the dissipative coupling can lead to a multimode coherent phonon source, opening a new avenue to study the phonon lasing in an anti-PT-Hamiltonian system and indicating that dissipation could be a useful resource for manipulating the interactions in an open complex system.

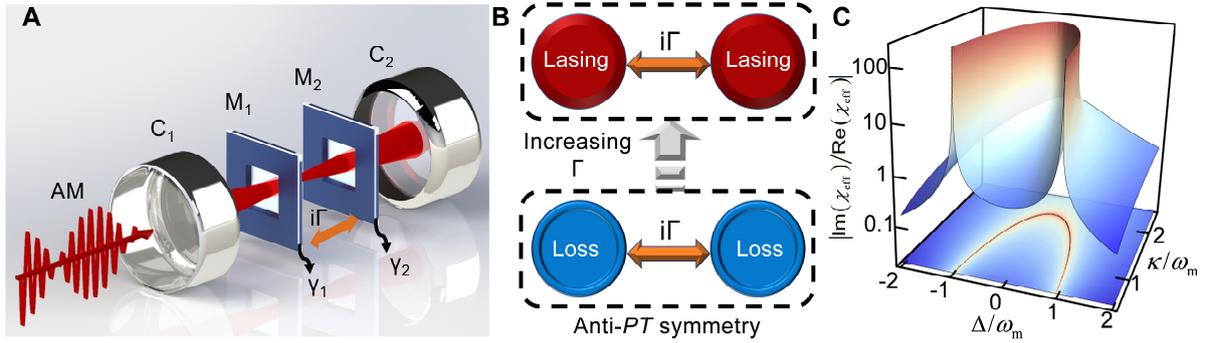

**Fig. 1. Dissipatively coupled optomechanical system.** (A) Schematic diagram of the two-membrane-in-the-middle optomechanical system. $C_{1,2}$, cavity mirrors; $M_{1,2}$, membranes; $\gamma_{1,2}$, mechanical decay rates; $i\Gamma$, purely dissipative coupling; AM, amplitude modulation. (B) The equivalent model of two dissipatively coupled oscillators. Increasing $\Gamma$ in such an anti-PT system leads to a multimode phonon laser. (C) The simulation of $|\text{Im}(\chi_{eff})/\text{Re}(\chi_{eff})|$ as functions of $\Delta/\omega_m$ and $\kappa/\omega_m$, where $\kappa$ is the cavity decay rate and $\omega_m$ is the mechanical frequency. $|\text{Im}(\chi_{eff})/\text{Re}(\chi_{eff})|$ can be infinitely large at $\Delta = \pm\sqrt{\omega_m^2 - \kappa^2/4} \approx \pm\omega_m$ in the resolved-sideband regime ($\kappa \ll \omega_m$) and exists a singularity at $\Delta=0$ and $\kappa=2\omega_m$.

Our experiment is performed in a two-membrane-in-the-middle optomechanical system, as shown in Fig. 1A, which is similar to the one used in Ref. [30]. Due to the inherent leakage of the cavity, there always exists a dissipative channel for the cavity field. However, it doesn't guarantee the purely dissipative coupling between two nanoresonators. In order to realize the purely dissipative coupling and clearly demonstrate the dissipative coupling induced phonon lasing, three steps are crucial. Firstly, the natural frequencies of two resonators are initially tailored to be different with $\omega_m/2\pi$=386 and 398 kHz (m=1,2) and the effective mechanical decays of the resonators are sophisticatedly engineered. Second, the cavity input field is amplitude modulated with the modulation frequency ω_d matching the frequency difference of resonators for the purpose of controlling the dissipative coupling strength independently. Third, the laser detuning Δ with respect to the optical cavity resonance has to be precisely tuned and the optomechanical system is in the resolved-sideband regime. Then, the system can be effectively described by a non-Hermitian Hamiltonian (see Supplementary Information, as shown in Fig. 1B)

$$\hat{H}_{eff} = \begin{pmatrix} \delta_d/2 - i\gamma_1'/2 & i\Gamma \\ i\Gamma & -\delta_d/2 - i\gamma_2'/2 \end{pmatrix}.$$

Here $\Gamma = g_1 g_2 \operatorname{Im}(\chi_{eff}) M/2$ represents the dissipative coupling strength, $\delta_d = \omega_d - (\omega_2 - \omega_1)$ is the frequency mismatch, and $\gamma_i' = \gamma_i - 2g_i^2 \operatorname{Im}(\chi_{eff})$ is the effective mechanical decay including the dynamic backaction. $\gamma_i$ is the intrinsic mechanical decay, $g_i$ is the optomechanical coupling strength for each membrane, and M is the modulation depth. $\chi_{eff}$ is the effective susceptibility introduced by the intracavity field. In the experiment, the optomechanical system is in the resolved-sideband regime, and $\Delta \approx \omega_m$, therefore the effective susceptibility is dominantly imaginary as illuminated in Fig. 1C. The complex eigenfrequencies of the hybridized modes for such a dissipatively coupled system are given by $\tilde{\omega}_\pm = -i(\gamma_1' + \gamma_2')/4 \pm \sqrt{\delta_d^2/4 - \Gamma^2 - (\gamma_1' - \gamma_2')^2/16 - i\delta_d(\gamma_1' - \gamma_2')/4} \equiv \omega_\pm - i\gamma_\pm/2$. As one can see that the dissipative coupling (iΓ) is essential for the onset of lasing, which flips the sign of decay rates of the coupled system (γ±). The lasing cannot happen with a coherent coupling.

Such a system provides the advantages for manipulating and controlling the critical experimental parameters. The dissipative coupling strength Γ can be controlled by tuning the modulation depth M without modifying other coefficients. During the experiment, the cavity input power $P_0$ is utilized to control the magnitude of $\chi_{eff}$ and Δ to tune the ratio of the imaginary and real parts of $\chi_{eff}$. $\gamma_i$ can also be modified by changing the vacuum pressure, and $\gamma_i'$ can be engineered to be the same or different by controlling $g_1$ and $g_2$, which is essential to observe the phenomena of level attraction. Therefore, such a two-membrane-in-the-middle optomechanical system provides an ideal platform for investigating a large range of interesting phenomena in non-Hermitian physics.

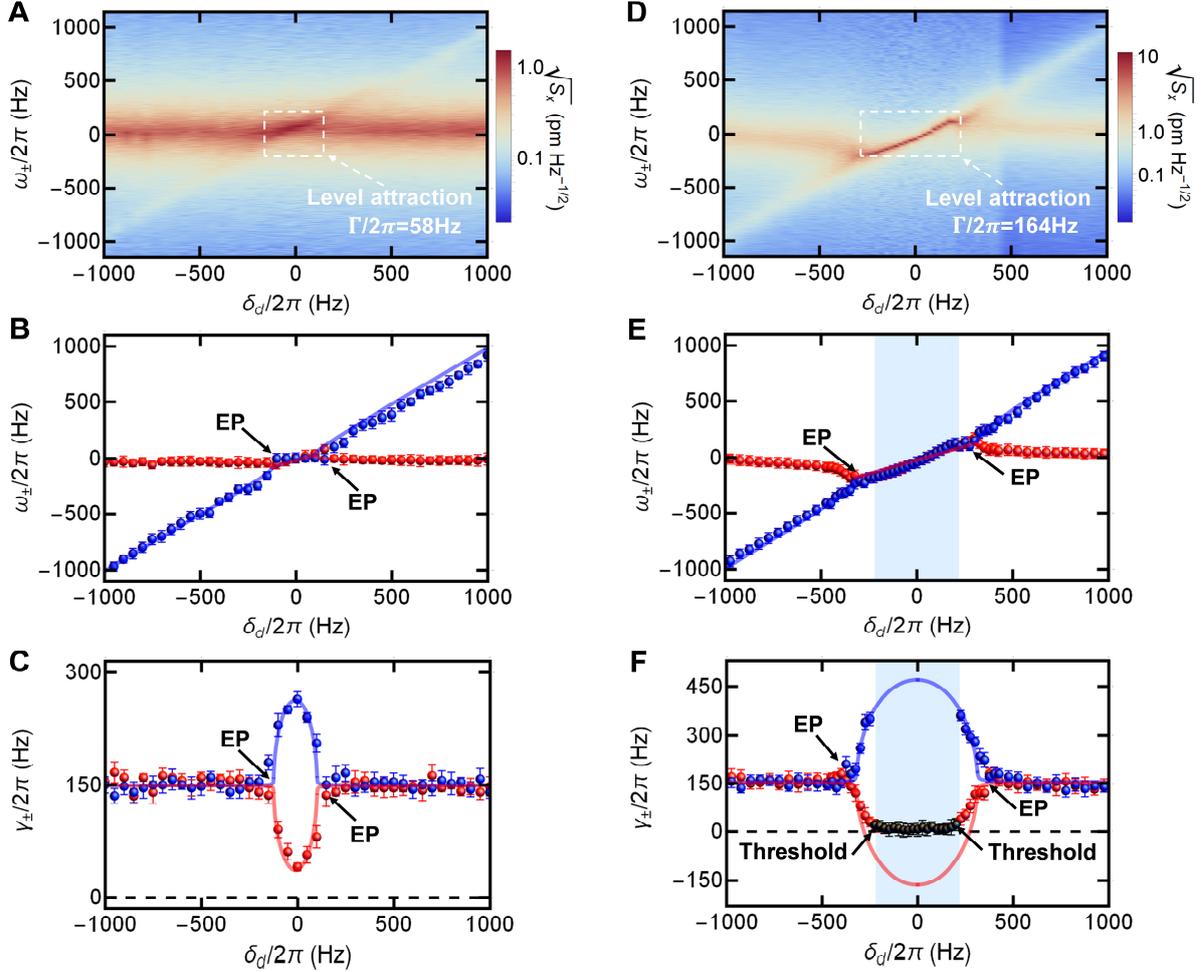

**Fig. 2. Level attraction and damping repulsion.** The measured mechanical power spectral density of the dissipatively coupled system as a function of $\delta_d$ for different coupling strengths (A) $\Gamma/2\pi=58$ Hz and (D) $\Gamma/2\pi=164$ Hz. (B,E) and (C,F) are the corresponding real and imaginary components, respectively. The dots are the experimental measurements and the solid curves are the theoretical simulations. The light blue region in (E) and (F) indicates the phonon lasing regime. The effective mechanical linewidths are $\gamma_1'/2\pi=\gamma_2'/2\pi= 150$ Hz.

The power spectral density of mechanical resonators is first examined by tuning $\gamma_1'=\gamma_2'=\gamma'$. The level attraction and damping repulsion are clearly observed, as shown in Fig. 2. As opposed to the anti-crossing phenomenon in a strongly coherently coupled system [31], the energy levels (the real components of the eigenfrequencies) of the coupled system merge in the region as the frequency mismatch is smaller than the dissipative coupling strength, i.e., $|\delta_d|\leq 2|\Gamma|$ (see Figs. 2B and 2E). On the contrary, the linewidths (the imaginary components of the eigenfrequencies) diverge, with one of the supermodes obtaining gain and the other one additional loss (see Figs. 2C and 2F). The EP locates at $\delta_d=\pm 2\Gamma$. Figures 2A-2C are measured at a relatively small dissipative coupling strength ($\Gamma/2\pi$=58 Hz), and no phonon lasing is observed, since the imaginary components are both positive even at the maximum hybridization, i.e., $\delta_d$=0. By increasing the dissipative coupling strength with a larger coupling strength ($\Gamma/2\pi$=164 Hz), as shown in Figs. 2D-2F, the region of level attraction is greatly broadened, and more importantly, the linewidth of gain mode becomes negative. The phonon lasing occurs after the threshold ($\delta_d = \pm 2\sqrt{\Gamma^2 - \gamma'^2/4}$) where the linewidth of gain mode becomes zero. The blue and red dots are the real and imaginary components of two eigenmodes extracted from Figs. 2A and 2D using the superposition of two Lorentz-dispersive lineshapes (see Supplementary Information), which agree well with the theoretical predictions. Since the noise power spectrum is measured in Figs. 2A and 2D, rather than the response spectrum as used in the cavity magnonics system [32], the fitting linewidth of the gain mode approaches to zero instead of a negative value in the phonon lasing regime, which is shown by the black dots in Fig. 2F (the Lorentz function is used for the fitting). The membranes are hybridized due to the dissipative coupling, and both membranes contain the gain mode. Therefore, both membranes are excited into the self-sustained regime simultaneously once the system passes the lasing threshold, and no mode competition is observed [13]. This is in contrast to the previous studies with the directly coherent pumping, where the phonon lasers are generated in sequence [15].

Figure 3 presents the mechanical spectral density as a function of $\Gamma$ at $\delta_d/2\pi$=250 Hz. When $\Gamma$ is relatively small, the real components are distinguishable and the imaginary components are degenerate. As $\Gamma$ increases, once the system exceeds the EP, the real components coalesce, and the imaginary components bifurcate, at which the eigenmodes also coalesce. Many extraordinary behaviors have been found near the EP, which is a singularity leading to the PT or anti-PT symmetry breaking transition [33-37]. By further increasing $\Gamma$, the linewidth of the gain mode (the blue dots in Fig. 3c) becomes smaller accompany with a higher oscillation amplitude, while the linewidth of the mode with additional loss (the red dots) becomes larger with a diminishing amplitude. Once the linewidth of the gain mode turns to be negative, the phonon lasing occurs, similar to Fig. 2F as $\delta_d$ varies. This lasing threshold is also a singularity, which is a bound state in the continuum and the group velocity can be zero [32].

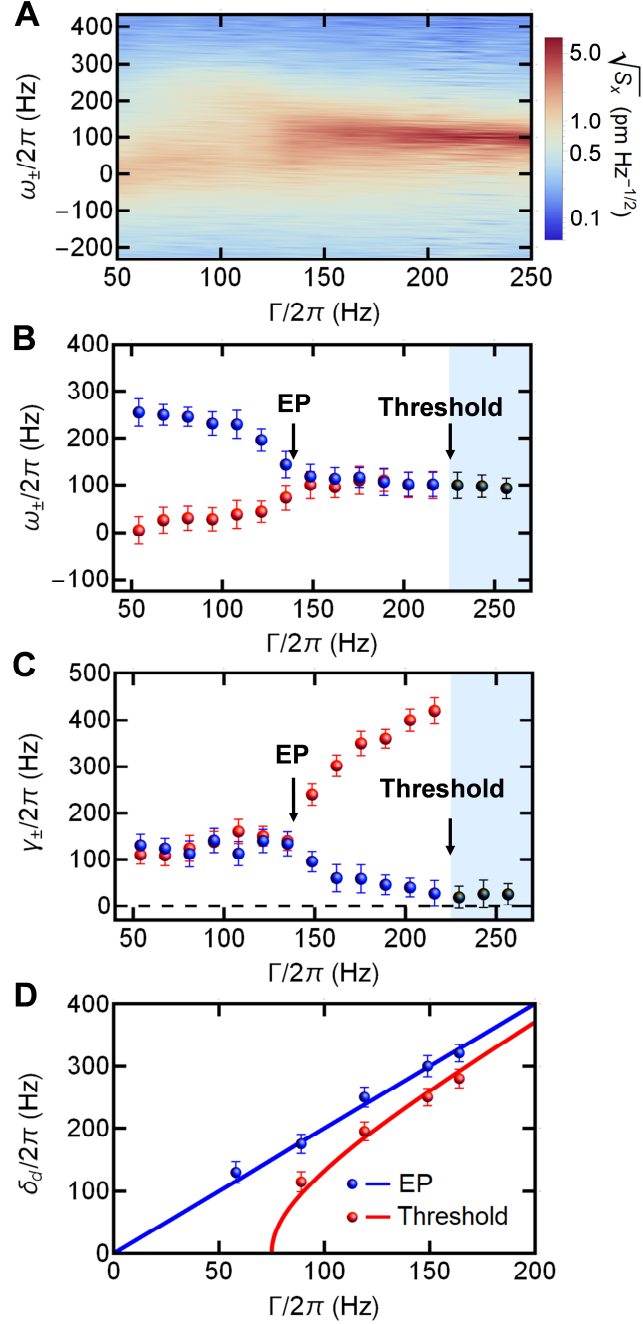

**Fig. 3. Exceptional point and lasing threshold.** (A) The measured mechanical power spectral density as a function of dissipative coupling strength Γ at $\delta_d/2\pi$=250 Hz. $\gamma'_1/2\pi=\gamma'_2/2\pi=$ 120 Hz. (B) and (C) are the corresponding real and imaginary components, respectively. The blue and red dots show the extracted real and imaginary components of two eigenmodes in the nonlasing regime, and the black dots denote the lasing mode. The light blue region in (B) and (C) indicates the phonon lasing regime. (D) Tunability of the EP and phonon lasing threshold. The dots are the experimental data and the curves are the theoretical simulation. $\gamma'_1/2\pi=\gamma'_2/2\pi=$ 150 Hz.

It is worth to point out that the EP and phonon lasing threshold are generally not coincident due to finite decays of the resonators. However, they are tunable by the frequency mismatch and dissipative coupling strength, as shown in Fig. 3D. The EP and lasing threshold are given by $\delta_d=\pm 2\Gamma$ and $\delta_d=\pm 2\sqrt{\Gamma^2-\gamma'^2/4}$ for $\gamma_1'=\gamma_2'=\gamma'$, respectively. They could be tuned to be the same at two situations: (1) The EP and lasing threshold approach to each other at a large $\Gamma$, and (2) $\gamma'=0$. However, the second case should be avoided here, since the phonon lasing is already generated at $\gamma'=0$ even $\Gamma=0$, which is attributed to the directly coherent pumping rather than the dissipative coupling induced lasing. When $\Gamma<\gamma'/2$, it is impossible to generate the phonon lasing.

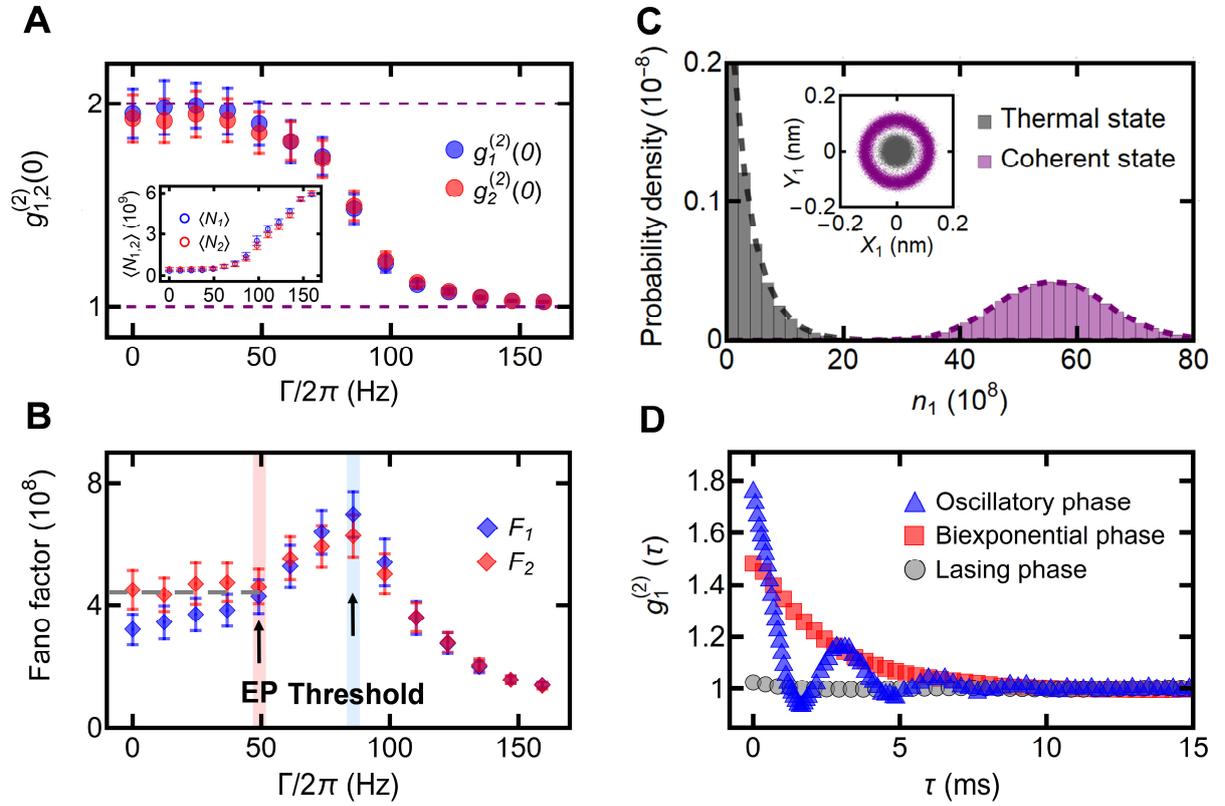

**Fig. 4. Phonon correlations and statistics.** (A,B) The second-order phonon correlation function at zero-time delay ($g_{1,2}^{(2)}(0)$) and Fano factor of both resonators as a function of $\Gamma$. The inset in Fig. 4A shows the corresponding phonon number ($\langle N_{1,2}\rangle$) as a function of $\Gamma$. $\delta_d/2\pi=100$ Hz and $\gamma_1'/2\pi=\gamma_2'/2\pi=120$ Hz. Each dot in Figs. 4A and 4B is the statistics over a time interval of 15 s (220000 data points). The grey dashed line in Fig. 4B is provided to guide the eye. (C) The measured phonon statistics at $\Gamma/2\pi=12$ Hz (grey) and 150 Hz (purple). The inset is the phase plot of the quadratures of the resonator's motion. (D) The second-order correlation as a function of time delay ($g_{1,2}^{(2)}(\tau)$) at $\Gamma/2\pi$=10 Hz, 80 Hz, and 150 Hz, which represent the oscillatory and biexponential phases in the nonlasing regime, and the coherence in the lasing regime, respectively. $\delta_d/2\pi$=300 Hz and $\gamma_1'/2\pi=\gamma_2'/2\pi$= 50 Hz.

We study phonon statistics and the second-order phonon correlation function in such a system. To do this, the displacement of resonator $x_{1,2}$ is measured in real time. Figure 4A shows the second-order phonon correlation function at zero-time delay ($g_{1,2}^{(2)}(0)=(\langle N_{1,2}^2\rangle-\langle N_{1,2}\rangle)/\langle N_{1,2}\rangle^2$) as a function of $\Gamma$, where $N_{1,2}\equiv\hat{b}_{1,2}^\dagger\hat{b}_{1,2}$ is the phonon number and $\langle N_{1,2}^2\rangle$ is the second moment of the distribution. It is clear that $g_{1,2}^{(2)}(0)$ is close to 2 initially, showing a property of thermal state. As $\Gamma$ increases, $g_{1,2}^{(2)}(0)$ decreases rapidly from 2 to 1 (the coherent state) near the phonon lasing threshold, which clearly indicates the onset of phonon lasing. The measured phonon numbers as a function of $\Gamma$ are plotted in the inset of Fig. 4A. The phonon number is obtained from the measurements via the relation $\langle N_{1,2}\rangle=m\omega_{1,2}\langle x_{1,2}^2\rangle/\hbar$, where $m$ is the mass and $\hbar$ is the reduced Planck constant. $\langle N_{1,2}\rangle$ displaces a threshold behavior as $\Gamma$ increases, which is the same as the conventional single mode phonon lasers. Since the origin of such a multimode phonon laser is the excitation of the coupled modes, $\langle N_1\rangle$ and $\langle N_2\rangle$ are identical when $\gamma_1'=\gamma_2'$. Figure 4C shows the phonon statistics of resonator for below ($\Gamma/2\pi$=12 Hz) and above ($\Gamma/2\pi$=150 Hz) the lasing threshold. The phonon statistics exhibits a thermal Brownian distribution and a Gaussian distribution, respectively, as expected for the thermal and phonon lasing states, which is also consistent with the results found in a levitated mesoscopic phonon laser [8].

To show the different phases for such a dissipative coupling induced phonon laser, the Fano factor (F=$(\Delta n)^2/\langle n\rangle$=1+$\langle n\rangle$g$^{(2)}$(0)-1) [5] is presented in Fig. 4B, in which the EP and lasing threshold are clearly observed. At $\Gamma/2\pi$=50 Hz, one can see a kink, which indicates the EP. This is because that the decay rates of the coupled system are degenerate and remain constant before the EP, which leads to a constant $\langle N_{1,2}\rangle$ and $g_{1,2}^{(2)}(0)$. After the EP, $\langle N_{1,2}\rangle$ and $g_{1,2}^{(2)}(0)$ start to increase and decrease, respectively. At $\Gamma/2\pi$=80 Hz, one can see a maximum, which determines the phonon lasing threshold [5].

The second-order correlation as a function of time delay ($g_{1,2}^{(2)}(\tau)$) reveals different phase transitions, as shown in Fig. 4D. Two different phases are identified in the nonlasing regime, which is in distinct contrast to the conventional phonon lasers. The blue triangles and red squares in Fig. 4D present $g_1^{(2)}(\tau)$ at $\Gamma/2\pi$=10 Hz and 80 Hz, which show the oscillatory (before the EP) and biexponential behaviors (between the EP and lasing threshold), respectively. Such a non-Hermitian phase transition is attributed to the EP, which locates at $\Gamma/2\pi$=50 Hz and separates the two phases. The origin of the oscillatory phase is due to the fact that the real components of the eigenvalues are nondegenerate and the imaginary components are degenerate before the EP, while it is just the reverse for the biexponential phase after the EP. This reveals that the non-Hermitian phase transition between the oscillatory and biexponential phases exists ubiquitously in dissipative interacting systems [38]. We use the fitting function $g_1^{(2)}(\tau)=C_1 e^{-\gamma_+ \tau}+C_2 e^{-\gamma_- \tau}+1$ to extract the decay rates in the biexponential phase [38], which are $\gamma_+/2\pi$=40 Hz and $\gamma_-/2\pi$=60 Hz. Above the lasing threshold, the coherence is further confirmed by $g_1^{(2)}(\tau) \sim 1$ at $\Gamma/2\pi$=150 Hz (the grey circles).

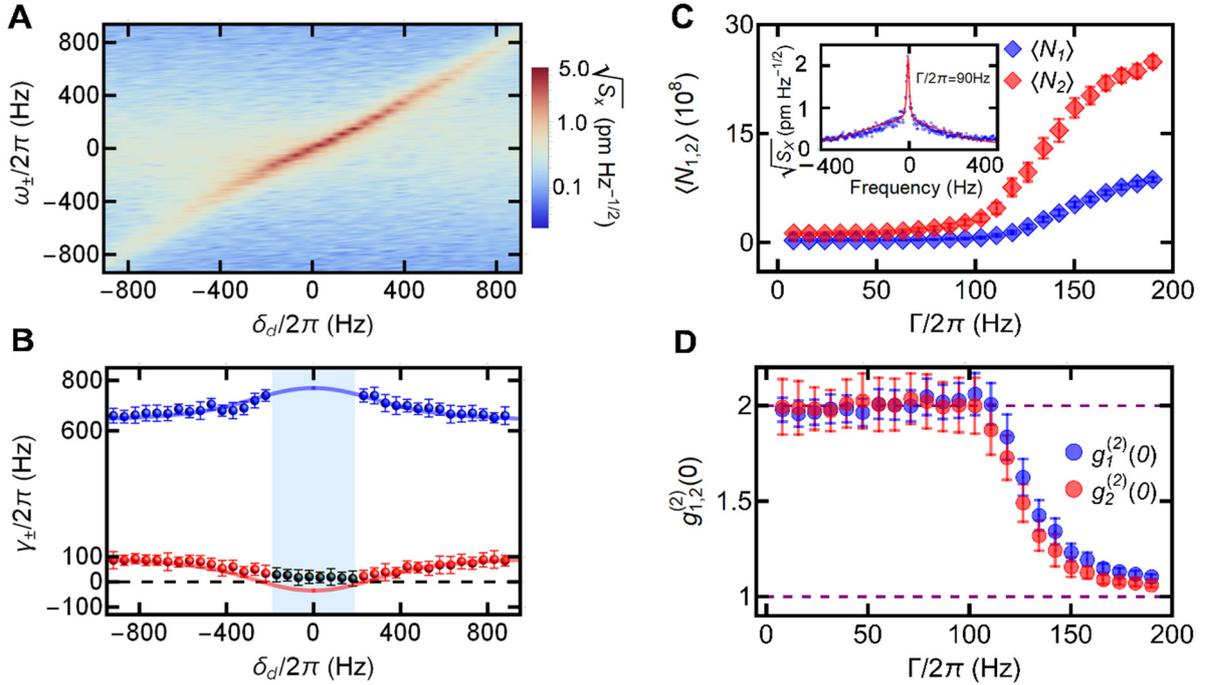

**Fig. 5. Electromagnetically induced absorption (EIA) with imbalanced decay rates.** (A) The measured mechanical power spectral density as a function of $\delta_d$ at $\Gamma/2\pi$=150 Hz. (B) The corresponding imaginary components. The light blue region indicates the phonon lasing regime. (C,D) $\langle N_{1,2} \rangle$ and $g_{1,2}^{(2)}(0)$ as a function of $\Gamma$ with $\delta_d$=0 Hz. The inset in Fig. 5C is the mechanical spectrum displacing the phenomenon analogous to EIA. $\gamma_1'/2\pi$= 650 Hz and $\gamma_2'/2\pi$= 100 Hz.

Finally, we investigate the situation of strongly imbalanced decay rates, i.e., $\gamma_1' \gg \gamma_2'$. In contrary to the case of $\gamma_1'=\gamma_2'$, no level attraction and EP are observed (see Figs. 5A and 5B), while a phenomenon analogous to EIA is observed in the phononic mode, as shown in the inset of Fig. 5C. The corresponding phenomenon in the coherent coupling regime is the so-called electromagnetically induced transparency (EIT) [39], which has been realized in the electromechanical systems known as the mechanically induced transparency [40]. The damping repulsion and multimode phonon lasing are still possible at a relatively small $\delta_d$ when the dissipative coupling is strong enough, i.e. one of the eigenmodes can receive gain and the linewidth can still be smaller than zero, as shown in Fig. 5B. Both membranes enter into the self-sustained regime simultaneously due to the hybridized nature of the system, which is similar to the situation of $\gamma_1'=\gamma_2'$. The difference is that the phonon numbers are distinct, as displaced in Fig. 5C, since the weights of the gain mode in each membrane are different. $g_{1,2}^{(2)}(0)$ as a function of $\Gamma$ is shown in Fig. 5D. Despite the significant difference of the phonon numbers, $g_1^{(2)}(0)$ and $g_2^{(2)}(0)$ are similar at different values of $\Gamma$. This indicates that both membranes have the same lasing threshold and are excited from thermal states to coherent states in a similar manner, in spite of the strongly imbalanced decay rates.

In conclusion, we have demonstrated that the dissipative coupling can lead to a multimode phonon laser in a two-membrane-in-the-middle optomechanical system. The level attraction and EIA have been observed in the phononic modes, which reveals that the coupling between two nanoresonators can be purely dissipative. The second-order correlation distinguishes the three different phases of the oscillatory, biexponential and coherent lasing behaviors. Our results may pave a new route to investigate non-Hermitian phononics with high flexibility and controllability in such the two-membrane-in-the-middle system.

[1] K. Vahala, M. Herrmann, S. Knünz, V. Batteiger, G. Saathoff, T. W. Hänsch, and Th. Udem, A phonon laser, Nat. Phys. 5, 682–686 (2009).

[2] M. Aspelmeyer, T. J. Kippenberg, and F. Marquardt, Cavity optomechanics, Rev. Mod. Phys. 86, 1391 (2014).

[3] I. S. Grudinin, H. Lee, O. Painter, and K. J. Vahala, Phonon Laser Action in a Tunable Two-Level System, Phys. Rev. Lett. 104, 083901 (2010).

[4] I. Mahboob, K. Nishiguchi, A. Fujiwara, and H. Yamaguchi, Phonon Lasing in an Electromechanical Resonator, Phys. Rev. Lett. 110, 127202 (2013).

[5] J. D. Cohen, S. M. Meenehan, G. S. MacCabe, S. Gröblacher, A. H. Safavi-Naeini, F. Marsili, M. D. Shaw, and O. Painter, Phonon counting and intensity interferometry of a nanomechanical resonator, Nature 520, 522 (2015).

[6] D. Navarro-Urrios, J. Gomis-Bresco, F. Alzina, N. E. Capuj, P. D. García, M. F. Colombano, E. Chavez-Angel, and C. M. Sotomayor-Torres, Self-sustained coherent phonon generation in optomechanical cavities, J. Opt. 18, 094006 (2016).

[7] J. Zhang, B. Peng, Ş. K. Özdemir, K. Pichler, D. O. Krimer, G. Zhao, F. Nori, Y. X. Liu, S. Rotter, and L. Yang, A phonon laser operating at an exceptional point, Nat. Photonics 12, 479 (2018).

[8] R. M. Pettit, W. Ge, P. Kumar, D. R. Luntz-Martin, J. T. Schultz, L. P. Neukirch, M. Bhattacharya, and A. Nick Vamivakas, An optical tweezer phonon laser, Nat. Photonics 13, 402 (2019).

[9] Y. Wen, N. Ares, F. J. Schupp, T. Pei, G. A. D. Briggs, and E. A. Laird, A coherent nanomechanical oscillator driven by single-electron tunneling, Nat. Phys. 16, 75 (2020).

[10] M. Zhang, G. S. Wiederhecker, S. Manipatruni, A. Barnard, P. McEuen, and M. Lipson, Synchronization of micromechanical oscillators using light, Phys. Rev. Lett. 109, 233906 (2012).

[11] M. Bagheri, M. Poot, L. Fan, F. Marquardt, and H. X. Tang, Photonic cavity synchronization of nanomechanical oscillators, Phys. Rev. Lett. 111, 213902 (2013).

[12] Matthew H. Matheny, Matt Grau, Luis G. Villanueva, Rassul B. Karabalin, M. C. Cross, and Michael L. Roukes, Phase Synchronization of Two Anharmonic Nanomechanical Oscillators, Phys. Rev. Lett. 112, 014101 (2014).

[13] U. Kemiktarak, M. Durand, M. Metcalfe, and J. Lawall, Mode competition and anomalous cooling in a multimode phonon laser, Phys. Rev. Lett. 113, 030802 (2014).

[14] E. Gil-Santos, M. Labousse, C. Baker, A. Goetschy, W. Hease, C. Gomez, A. Lemaitre, G. Leo, C. Ciuti, and I. Favero, Light-mediated cascaded locking of multiple nano-optomechanical oscillators, Phys. Rev. Lett. 118, 063605 (2017).

[15] Jiteng Sheng, Xinrui Wei, Cheng Yang, and Haibin Wu, Self-Organized Synchronization of Phonon Lasers, Phys. Rev. Lett. 124, 053604 (2020).

[16] Laura Mercadé, Karl Pelka, Roel Burgwal, André Xuereb, Alejandro Martínez, and Ewold Verhagen, Floquet phonon lasing in multimode optomechanical systems, Phys. Rev. Lett. 127, 073601 (2021).

[17] X. L. Feng, C. J. White, A. Hajimiri, and M. L. Roukes, A self-sustaining ultrahigh-frequency nanoelectromechanical oscillator, Nat. Nanotechnol. 3, 342 (2008).

[18] Kejie Fang, Matthew H. Matheny, Xingsheng Luan, and Oskar Painter, Optical transduction and routing of microwave phonons in cavity-optomechanical circuits, Nat. Photonics 10, 489 (2016).

[19] Nishant Dogra, Manuele Landini, Katrin Kroeger, Lorenz Hruby, Tobias Donner, Tilman Esslinger, Dissipation-induced structural instability and chiral dynamics in a quantum gas, Science 366, 1496 (2019).

[20] Youngsun Choi, Choloong Hahn, Jae Woong Yoon, and Seok Ho Song, Observation of an anti-PT-symmetric exceptional point and energy-difference conserving dynamics in electrical circuit resonators, Nat. Commun. 9, 2182 (2018).

[21] X. Wei, J. Sheng, C. Yang, Y. Wu, and H. Wu, Controllable two-membrane-in-the-middle cavity optomechanical system, Phys. Rev. A 99, 023851 (2019).

[22] P. Peng, W. Cao, C. Shen, W. Qu, J. Wen, L. Jiang, and Y. Xiao, Anti-parity–time symmetry with flying atoms, Nat. Phys. 12, 1139 (2016).

[23] Fan Yang, Yong-Chun Liu, and Li You, Anti-PT symmetry in dissipatively coupled optical systems, Phys. Rev. A 96, 053845 (2017).

[24] Y. Li, Y.-G. Peng, L. Han, M.-A. Miri, W. Li, M. Xiao, X.-F. Zhu, J. Zhao, A. Alù, S. Fan, and C.-W. Qiu, Anti-Parity-Time Symmetry in Diffusive Systems, Science 364, 170 (2019).

[25] Zhao-Hui Peng, Chun-Xia Jia, Yu-Qing Zhang, Ji-Bing Yuan, and Le-Man Kuang, Level attraction and PT symmetry in indirectly coupled microresonators, Phys. Rev. A 102, 043527 (2020).

[26] M. Harder, Y. Yang, B. M. Yao, C. H. Yu, J. W. Rao, Y. S. Gui, R. L. Stamps, and C.-M. Hu, Level Attraction Due to Dissipative Magnon-Photon Coupling, Phys. Rev. Lett. 121, 137203 (2018).

[27] N. R. Bernier, L. D. Tóth, A. K. Feofanov, and T. J. Kippenberg, Level attraction in a microwave optomechanical circuit, Phys. Rev. A 98, 023841 (2018).

[28] T. M. Karg, B. Gouraud, C. T. Ngai, G.-L. Schmid, K. Hammerer, and P. Treutlein, Light-mediated strong coupling between a mechanical oscillator and atomic spins 1 meter apart, Science 369, 174 (2020).

[29] Xihua Yang, Jiteng Sheng, and Min Xiao, Electromagnetically induced absorption via incoherent collisions, Phys. Rev. A 84, 043837 (2011).

[30] Cheng Yang, Xinrui Wei, Jiteng Sheng, and Haibin Wu, Phonon heat transport in cavity-mediated optomechanical nanoresonators, Nat. Commun. 11, 4656 (2020).

[31] Jiteng Sheng, Yuanxi Chao, Santosh Kumar, Haoquan Fan, Jonathon Sedlacek, and James P. Shaffer, Intracavity Rydberg-atom electromagnetically induced transparency using a high-finesse optical cavity, Phys. Rev. A 96, 033813 (2017).

[32] Y. Yang, Yi-Pu Wang, J. W. Rao, Y. S. Gui, B. M. Yao, W. Lu, and C.-M. Hu, Unconventional Singularity in Anti-Parity-Time Symmetric Cavity Magnonics, Phys. Rev. Lett. 125, 147202 (2020).

[33] M. Bender and S. Boettcher, Real Spectra in Non-Hermitian Hamiltonians Having PT Symmetry, Phys. Rev. Lett. 80, 5243 (1998).

[34] Jiteng Sheng, Mohammad-Ali Miri, Demetrios N. Christodoulides, and Min Xiao, PT-symmetric optical potentials in a coherent atomic medium, Phys. Rev. A 88, 041803(R) (2013).

[35] R. El-Ganainy, K. G. Makris, M. Khajavikhan, Z. H. Musslimani, S. Rotter, and D. N. Christodoulides, Non-Hermitian physics and PT symmetry, Nat. Phys. 14, 11 (2018).

[36] Mohammad-Ali Miri and Andrea Alù, Exceptional points in optics and photonics, Science 363, eaar7709 (2019).

[37] Ş. K. Özdemir, S. Rotter, F. Nori, and L. Yang, Parity–time symmetry and exceptional points in photonics, Nat. Materials 18, 783 (2019).

[38] F. E. Öztürk, T. Lappe, G. Hellmann, J. Schmitt, J. Klaers, F. Vewinger, J. Kroha, M. Weitz, Observation of a non-Hermitian phase transition in an optical quantum gas, Science 372, 88 (2021).

[39] Michael Fleischhauer, Atac Imamoglu, and Jonathan P. Marangos, Electromagnetically induced transparency: Optics in coherent media, Rev. Mod. Phys. 77, 633 (2005).

[40] I. Mahboob, K. Nishiguchi, H. Okamoto, and H. Yamaguchi, Phonon-cavity electromechanics, Nat. Phys. 8, 387 (2012).

# Supplementary Information

**Supplementary Note 1: Effective Hamiltonian for dissipative coupling**

The dissipative interaction is based on the indirect coupling between two spatially separated nanomechanical resonators mediated by a common optical field in a two-membrane-in-the-middle cavity optomechanical system. The cavity is driven by a laser field with amplitude modulation. The total Hamiltonian of such a system in the rotating frame of the driving laser frequency can be written as $(\hbar = 1)$ [1-3]

$$\hat{H} = -\Delta \hat{a}^\dagger \hat{a} + \omega_1 \hat{b}_1^\dagger \hat{b}_1 + \omega_2 \hat{b}_2^\dagger \hat{b}_2 - g_1 \hat{a}^\dagger \hat{a} \left( \hat{b}_1^\dagger + \hat{b}_1 \right) - g_2 \hat{a}^\dagger \hat{a} \left( \hat{b}_2^\dagger + \hat{b}_2 \right) + i\eta(t)\left( \hat{a}^\dagger - \hat{a} \right). \quad \text{(S1)}$$

Here $\hat{a}$ and $\hat{b}_{1,2}$ are the annihilation operators of the cavity mode and the mechanical oscillators, respectively. $\Delta = \omega_L - \omega_C$ is the frequency detuning between the driving laser and the cavity resonance. $\omega_{1,2}$ is the intrinsic frequency of the mechanical oscillator. $g_{1,2}$ is the optomechanical coupling strength. $\eta(t) = \sqrt{P(t)\kappa_{in}/\hbar\omega_L}$ is the driving strength. $P(t) = P_0(1 + M\cos\omega_d t)$ is the input laser power with a constant power $P_0$, modulation depth $M$, and modulation frequency $\omega_d$. $\kappa_{in}$ is the loss of the cavity input mirror.

According to Eq. (S1), the quantum Langevin equations can be written as follows

$$\partial_t \hat{a} = \left[ -\kappa/2 + i\Delta + i\sum_{j=1}^{2} g_j \left( \hat{b}_j + \hat{b}_j^\dagger \right) \right] \hat{a} + \eta(t) + \sqrt{\kappa}\hat{a}_{in}, \quad \text{(S2a)}$$

$$\partial_t \hat{b}_j = -\left( \gamma_j/2 + i\omega_j \right) \hat{b}_j + ig_j \hat{a}^\dagger \hat{a} + \sqrt{\gamma_j}\hat{b}_{jin}, \quad (j = 1, 2) \quad \text{(S2b)}$$

where $\kappa$ is the total cavity decay rate and $\gamma_j$ is the mechanical damping rate. $\hat{a}_{in}$ and $\hat{b}_{jin}$ are the Markovian quantum noise operators of cavity mode and $j^{th}$ mechanical mode, respectively. When the cavity is driven by a classical field, we can linearly decompose the cavity mode into the average field $\langle \hat{a} \rangle = \alpha$ and the corresponding fluctuation $\delta\hat{a}$, i.e., $\hat{a} \to \alpha + \delta\hat{a}$. When the modulation depth $M$ of the driven laser is much less than one, we can retain the first-order expansion of the driving strength $\eta(t) \approx \sqrt{P_0\kappa_{in}/\hbar\omega_L}\left(1+e^{i\omega_d t}M/4+e^{-i\omega_d t}M/4\right)$. Then, the average intracavity amplitude can be obtained as $\alpha(t)=\alpha_0+\alpha_{-1}e^{i\omega_d t}+\alpha_1 e^{-i\omega_d t}$, where $\alpha_0=\eta_0\chi_c(0)$, $\alpha_{-1}=\eta_0\chi_c(-\omega_d)M/4$, and $\alpha_1=\eta_0\chi_c(\omega_d)M/4$, with the susceptibility function of cavity field $\chi_c(\omega)=\left[\kappa/2-i(\Delta+\omega)\right]^{-1}$ and $\eta_0=\sqrt{P_0\kappa_{in}/\hbar\omega_L}$. By assuming $\omega_d \ll \Delta$, the intracavity amplitude is approximately as $\alpha(t)\approx\eta_0\chi_c(0)\left(1+e^{i\omega_d t}M/4+e^{-i\omega_d t}/4\right)$.

In order to obtain the explicit expression of the effective interaction between two mechanical modes, we utilize the transformations $\hat{b}_j \to \hat{b}_j e^{-i\Omega_j t}$ and $\hat{b}_{jin} \to \hat{b}_{jin} e^{-i\Omega_j t}$. Thus, the dynamics of intracavity field fluctuation $\delta\hat{a}$ and the mechanical mode $\hat{b}_j$ can be obtained as

$$\partial_t \delta\hat{a} = (-\kappa/2+i\Delta)\delta\hat{a} + i\alpha\sum_{j=1}^{2}g_j\left(\hat{b}_j e^{-i\Omega_j}+\hat{b}_j^\dagger e^{i\Omega_j}\right)+\sqrt{\kappa}\hat{a}_{in}, \qquad (S3a)$$

$$\partial_t \hat{b}_j = -\left(\gamma_j/2+i(\omega_j-\Omega_j)\right)\hat{b}_j + ig_j\left(\alpha^*\delta\hat{a}+\alpha\delta\hat{a}^\dagger\right)e^{i\Omega_j}+\sqrt{\gamma_j}\hat{b}_{jin}. \qquad (S3b)$$

Due to the fact that the cavity damping rate is much larger than the mechanical loss, the intracavity field adiabatically follows the dynamics of the mechanical modes [4,5]. From Eq. (S3a), we can obtain the formal solution of intracavity fluctuation on long time scales compared with $\kappa^{-1}$ as follows

$$\delta\hat{a}(t)=\sqrt{\kappa}\hat{f}_{in}(t)+i\sum_{j=1}^{2}g_j\left[A_j(\omega_d)\hat{b}_j e^{-i\Omega_j t}+B_j(\omega_d)\hat{b}_j^\dagger e^{i\Omega_j t}\right], \qquad (S4)$$

where

$$A_j(\omega_d)=\left[\alpha_0\chi_c(\Omega_j)+\alpha_{-1}\chi_c(\Omega_j-\omega_d)e^{i\omega_d t}+\alpha_1\chi_c(\Omega_j+\omega_d)e^{-i\omega_d t}\right],$$

$$B_j(\omega_d)=\left[\alpha_0\chi_c(-\Omega_j)+\alpha_{-1}\chi_c(-\Omega_j-\omega_d)e^{i\omega_d t}+\alpha_1\chi_c(-\Omega_j+\omega_d)e^{-i\omega_d t}\right],$$

and $\hat{f}_{in}(t)=\int_0^t d\tau e^{(i\Delta-\kappa/2)(t-\tau)}\hat{a}_{in}(\tau)$ is the noise operator.

By substituting Eq. (S4) into Eq. (S3b) and neglecting all the noise operator terms including $\hat{f}_{in}$ and $\hat{b}_{jin}$, we can obtain the time-dependent coupled-mode equations of mechanical oscillators

$$i\partial_t \hat{b}_1 = \left(-i\gamma_1/2 + (\omega_1 - \Omega_1)\right)\hat{b}_1 + \Sigma_{11}\hat{b}_1 + \Sigma_{12}\hat{b}_2 e^{-i(\Omega_2-\Omega_1)t}, \tag{S5a}$$

$$i\partial_t \hat{b}_2 = \Sigma_{21}\hat{b}_1 e^{i(\Omega_2-\Omega_1)t} + \left(-i\gamma_2/2 + (\omega_2 - \Omega_2)\right)\hat{b}_2 + \Sigma_{22}\hat{b}_2, \tag{S5b}$$

where $\Sigma_{jk} = -ig_j g_k \left(\alpha^* A_k - \alpha B_k^*\right)$ and

$$\left(\alpha^* A_k - \alpha B_k^*\right) =$$
$$\alpha_0^* \alpha_0 \left(\chi_c(\omega_k) - \chi_c^*(-\omega_k)\right) + \alpha_{-1}^* \alpha_{-1}\left(\chi_c(\omega_k - \omega_d) - \chi_c^*(-\omega_k - \omega_d)\right) + \alpha_1^* \alpha_1\left(\chi_c(\omega_k + \omega_d) - \chi_c^*(-\omega_k + \omega_d)\right)$$
$$+ e^{i\omega_d t}\left[\alpha_0^* \alpha_{-1}\left(\chi_c(\omega_k - \omega_d) - \chi_c^*(-\omega_k)\right) + \alpha_1^* \alpha_0\left(\chi_c(\omega_k) - \chi_c^*(-\omega_k + \omega_d)\right)\right]$$
$$+ e^{-i\omega_d t}\left[\alpha_0^* \alpha_1\left(\chi_c(\omega_k + \omega_d) - \chi_c^*(-\omega_k)\right) + \alpha_{-1}^* \alpha_0\left(\chi_c(\omega_k) - \chi_c^*(-\omega_k - \omega_d)\right)\right]$$
$$+ e^{i2\omega_d t}\alpha_1^* \alpha_{-1}\left(\chi_c(\omega_k - \omega_d) - \chi_c^*(-\omega_k + \omega_d)\right) + e^{-i2\omega_d t}\alpha_{-1}^* \alpha_1\left(\chi_c(\omega_k + \omega_d) - \chi_c^*(-\omega_k - \omega_d)\right).$$

Further, we drop the fast oscillating terms contained in $\Sigma_{11}$, $\Sigma_{22}$, $\Sigma_{12}e^{-i(\Omega_2-\Omega_1)t}$ and $\Sigma_{21}e^{i(\Omega_2-\Omega_1)t}$, and Eq. (S5) is simplified to be

$$i\partial_t \hat{b}_1 = \left(-i\gamma_1/2 + (\omega_1 - \Omega_1)\right)\hat{b}_1 + \sigma_{11}\hat{b}_1 + \sigma_{12}\hat{b}_2 e^{i\omega_d t}e^{-i(\Omega_2-\Omega_1)t}, \tag{S6a}$$

$$i\partial_t \hat{b}_2 = \sigma_{21}\hat{b}_1 e^{-i\omega_d t}e^{i(\Omega_2-\Omega_1)t} + \left(-i\gamma_2/2 + (\omega_2 - \Omega_2)\right)\hat{b}_2 + \sigma_{22}\hat{b}_2, \tag{S6b}$$

where

$$\sigma_{11} = -ig_1^2\left[\alpha_0^* \alpha_0\left(\chi_c(\omega_1) - \chi_c^*(-\omega_1)\right) + \alpha_{-1}^* \alpha_{-1}\left(\chi_c(\omega_1 - \omega_d) - \chi_c^*(-\omega_1 - \omega_d)\right) + \alpha_1^* \alpha_1\left(\chi_c(\omega_1 + \omega_d) - \chi_c^*(-\omega_1 + \omega_d)\right)\right],$$

$$\sigma_{22} = -ig_2^2\left[\alpha_0^* \alpha_0\left(\chi_c(\omega_2) - \chi_c^*(-\omega_2)\right) + \alpha_{-1}^* \alpha_{-1}\left(\chi_c(\omega_2 - \omega_d) - \chi_c^*(-\omega_2 - \omega_d)\right) + \alpha_1^* \alpha_1\left(\chi_c(\omega_2 + \omega_d) - \chi_c^*(-\omega_2 + \omega_d)\right)\right],$$

$$\sigma_{12} = -ig_1 g_2\left[\alpha_0^* \alpha_{-1}\left(\chi_c(\omega_2 - \omega_d) - \chi_c^*(-\omega_2)\right) + \alpha_1^* \alpha_0\left(\chi_c(\omega_2) - \chi_c^*(-\omega_2 + \omega_d)\right)\right],$$

$$\sigma_{21} = -ig_2 g_1\left[\alpha_0^* \alpha_1\left(\chi_c(\omega_1 + \omega_d) - \chi_c^*(-\omega_1)\right) + \alpha_{-1}^* \alpha_0\left(\chi_c(\omega_1) - \chi_c^*(-\omega_1 - \omega_d)\right)\right].$$

Thus, the effective time-dependent Hamiltonian of mechanical modes can be extracted from Eq. (S6) as follows

$$\hat{H}_D(t) = \begin{pmatrix} -i\gamma_1/2 + (\omega_1 - \Omega_1) + \sigma_{11} & \sigma_{12} e^{i(\omega_d - (\Omega_2 - \Omega_1))t} \\ \sigma_{21} e^{-i(\omega_d - (\Omega_2 - \Omega_1))t} & -i\gamma_2/2 + (\omega_2 - \Omega_2) + \sigma_{22} \end{pmatrix}. \quad (S7)$$

Using the unitary transformation

$$U(t) = \begin{pmatrix} e^{i\delta_d t/2} & 0 \\ 0 & e^{-i\delta_d t/2} \end{pmatrix}$$

with $\delta_d = \omega_d - (\Omega_2 - \Omega_1)$, $\hat{H}_D(t)$ can be transformed into the following time-independent Hamiltonian $\hat{H}_D^{rot} = U^{-1}\hat{H}_D(t)U - iU^{-1}\partial_t U$, i.e.,

$$\hat{H}_D^{rot} = \begin{pmatrix} \delta_d/2 + (\omega_1 - \Omega_1) - i\gamma_1/2 + \sigma_{11} & \sigma_{12} \\ \sigma_{21} & -\delta_d/2 + (\omega_2 - \Omega_2) - i\gamma_2/2 + \sigma_{22} \end{pmatrix}. \quad (S8)$$

By assuming $\omega_d \ll \omega_{1,2}$, we have $\chi_c(\omega_{1,2}) \approx \chi_c(\omega_{1,2} \pm \omega_d)$. Thus, $\sigma_{jk}$ can be simplified as $\sigma_{11} \approx g_1^2 \chi_{eff}$, $\sigma_{22} \approx g_2^2 \chi_{eff}$, and $\sigma_{12} \approx \sigma_{21} \approx g_1 g_2 \chi_{eff} M/2$. Here $\chi_{eff} = -i|\eta_0 \chi_c(0)|^2 (\chi_c(\omega_m) - \chi_c^*(-\omega_m))$ is effective susceptibility introduced by the intracavity field with $\omega_m = (\omega_1 + \omega_2)/2$, i.e.,

$$\chi_{eff} = \frac{P_0 \kappa_{in}/\hbar \omega_L}{\kappa^2/4 + \Delta^2} \left( \frac{-i\kappa/2 + (\Delta + \omega_m)}{\kappa^2/4 + (\Delta + \omega_m)^2} + \frac{i\kappa/2 + (\Delta - \omega_m)}{\kappa^2/4 + (\Delta - \omega_m)^2} \right), \quad (S9)$$

which results from the dynamical backaction with the optomechanical spring and damping effects. Thus, Eq. (S7) can be shown as

$$\hat{H}_D^{rot} = \begin{pmatrix} \delta_d/2 - i\gamma_1/2 + g_1^2 \chi_{eff} & g_1 g_2 \chi_{eff} M/2 \\ g_1 g_2 \chi_{eff} M/2 & -\delta_d/2 - i\gamma_2/2 + g_2^2 \chi_{eff} \end{pmatrix}, \quad (S10)$$

with complex off-diagonal elements $g_1 g_2 \chi_{eff} M/2$ describing the coherent (real part) and dissipative (imaginary part) interactions between two mechanical oscillators, which depends on $P_0$, $\Delta$, and $M$.

The real ($\text{Re}(\chi_{\text{eff}})$) and imaginary ($\text{Im}(\chi_{\text{eff}})$) parts of the effective susceptibility, and the absolute value of $\text{Im}(\chi_{\text{eff}})/\text{Re}(\chi_{\text{eff}})$ are plotted as a function of the cavity detuning $\Delta/\omega_m$ in both unresolved- and resolved-sideband regimes in Fig. S1. It is straightforward to see that the resolved-sideband regime is desired for the purely imaginary interaction.

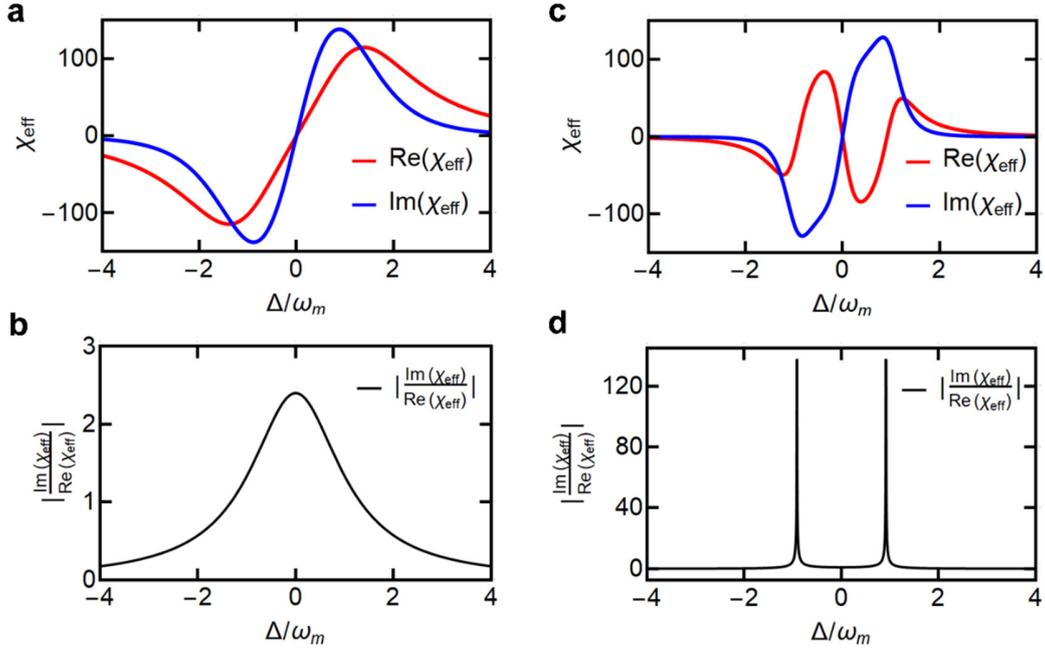

FIG. S1. The real and imaginary parts of $\chi_{\text{eff}}$, and the absolute value of $\text{Im}(\chi_{\text{eff}})/\text{Re}(\chi_{\text{eff}})$ as a function of cavity detuning $\Delta/\omega_m$. (a,b) The unresolved-sideband regime ($\kappa=3\omega_m$) and (c,d) the resolved-sideband regime ($\kappa=0.8\omega_m$).

In the resolved-sideband regime, $\chi_{\text{eff}}$ is dominantly imaginary if the cavity detuning is close to the mechanical frequency $(\Delta \approx \pm\omega_{1,2})$, as shown in Fig. S1d. Consequently, the effective interaction between two mechanical modes can be treated as purely dissipative. Moreover, if we let $\omega_1=\Omega_1$ and $\omega_2=\Omega_2$ in Eq. (S10), the effective two-mode Hamiltonian for dissipative coupling $\hat{H}_{\text{eff}}$ between two mechanical modes can be shown as

$$\hat{H}_{\text{eff}} = \begin{pmatrix} \delta_d/2 - i\gamma_1/2 + ig_1^2\,\text{Im}(\chi_{\text{eff}}) & ig_1g_2\,\text{Im}(\chi_{\text{eff}})M/2 \\ ig_2g_1\,\text{Im}(\chi_{\text{eff}})M/2 & -\delta_d/2 - i\gamma_2/2 + ig_2^2\,\text{Im}(\chi_{\text{eff}}) \end{pmatrix}. \tag{S11}$$

**Supplementary Note 2: Coexistence of dissipative and coherent interactions**

In the main text, we mainly discuss the purely dissipative interaction. Here, we briefly investigate the situation when the coherent interaction ( $g_1 g_2 \text{Re}(\chi_{eff}) M / 2$ ) between two membranes cannot be ignored, which is realized by controlling the cavity detuning.

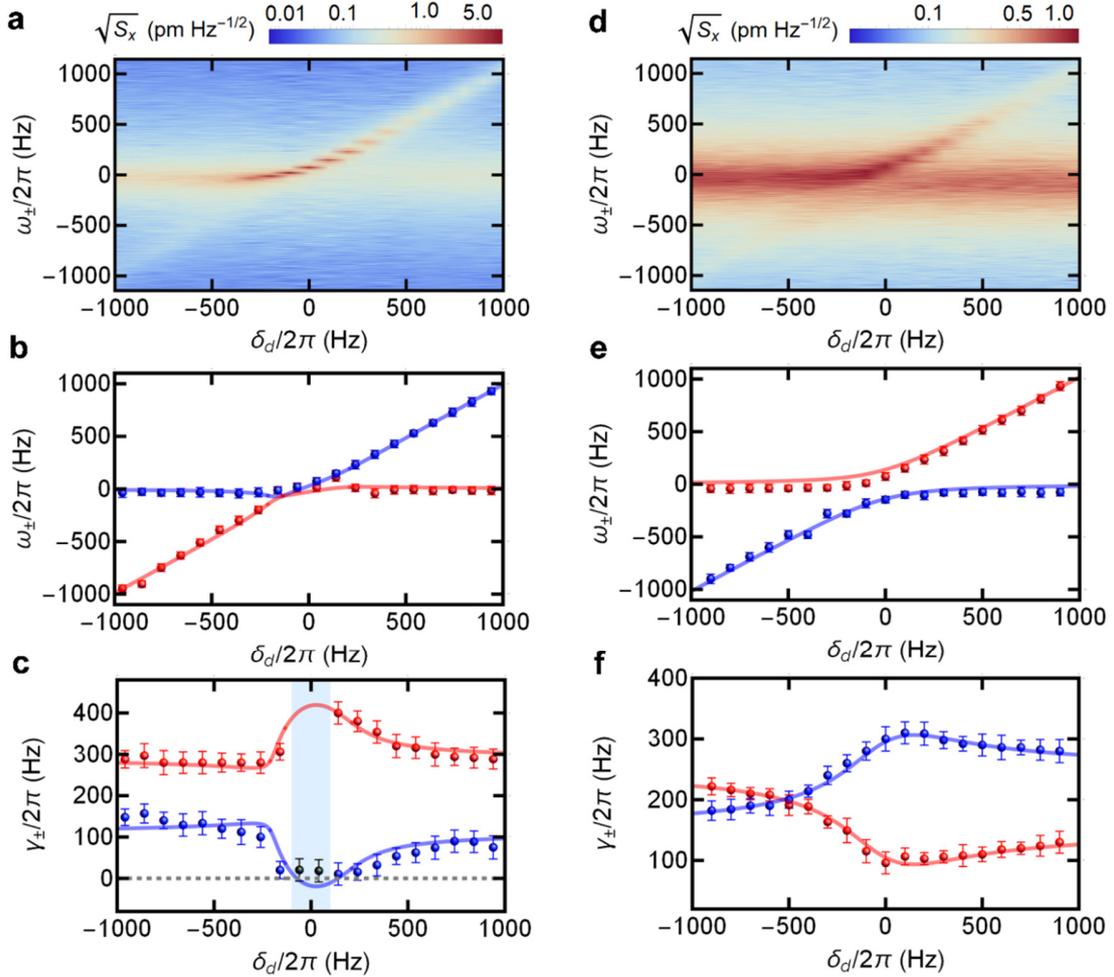

FIG. S2. The measured mechanical power spectral density of the dissipatively and coherently coupled system as a function of effective frequency detuning $\delta_d$ for the different ratios of (a) $|\text{Im}(\chi_{eff})/\text{Re}(\chi_{eff})| \approx 3$ and (d) $|\text{Im}(\chi_{eff})/\text{Re}(\chi_{eff})| \approx 0.3$. (b,e) and (c,f) are the corresponding eigenfrequencies $\omega_\pm$ and linewidths $\gamma_\pm$ of the normal mode, respectively. The dots are the experimental measurements and the solid curves are the theoretical simulations. The system

parameters are $\gamma_1'/2\pi \approx 110\text{Hz}$, $\gamma_2'/2\pi \approx 290\text{Hz}$, and $\Gamma/2\pi \approx 100\text{Hz}$ for (a-c), and $\gamma_1'/2\pi \approx 150\text{Hz}$, $\gamma_2'/2\pi \approx 250\text{Hz}$, and $\Gamma/2\pi \approx 47\text{Hz}$ for (d-f).

The mechanical power spectral densities of membrane are plotted in Fig. S2 as a function of $\delta_d$. The level attraction of normal modes can be observed when the coherent interaction is relatively smalle compared to the dissipative coupling strength with a ratio $|\text{Im}(\chi_{eff})/\text{Re}(\chi_{eff})| \approx 3$, as shown in Fig. S2a and Fig. S2b. The level repulsion of normal modes can be observed when the coherent interaction is relatively large with a ratio $|\text{Im}(\chi_{eff})/\text{Re}(\chi_{eff})| \approx 0.3$, as shown in Fig. S2d and Fig. S2e. The asymmetry of the linewidths with respect to $\delta_d$, as shown in Figs. S2c and S2f, is due to the unbalanced linewidths of bare modes and the non-negligible coherent coupling. The phonon lasing is also observed in spite of the mixed coherent interaction, as shown in Fig. S2c.

**Supplementary Note 3: Experimental setup**

The experiment is performed in an optomechanical system with two spatially separated membranes inside an optical cavity in the resolved-sideband regime [3]. The optical cavity consists of two mirrors with reflectivity ~ 0.9998 (at the wavelength of 1064nm) with a cavity length of 36 mm. Two stoichiometric silicon nitride membranes with a size $1\times 1\text{mm}^2$ and a thickness of 50nm are placed inside the optical cavity. The cavity finesse with two membranes in the middle is about 12000. We can precisely control the natural frequency and position of each membrane by the attached piezos. A narrow linewidth 1064nm Nd: YAG laser is split into two beams. One weak locking beam passes through an electro-optic modulator (EOM) to stabilize the cavity frequency via the Pound-Drever-Hall technique. The other strong driving beam goes through two-cascade acousto-optic modulators (AOM) to control the frequency detuning between the driving laser and optical cavity. The whole two-membrane-in-the-middle cavity optomechanical system is placed inside a vacuum chamber. The cavity driving field is amplitude modulated via the AOM. The mechanical motion of two membranes can be detected by a weak 795nm probe laser reflecting from two membranes (not resonant with the cavity). A spectrum analyzer and a lock-in amplifier are used for the measurements of membranes' motions.

**Supplementary Note 4: Data analysis**

In this note, we first describe the data fitting for extracting the eigenvalues from the measured mechanical spectrum. According to the effective Hamiltonian, the noise power spectrum of each phonon mode can be obtained as follows

$$S_{n_1}(\omega) = \frac{\left[\gamma_2^{'2}/4 + (\omega+\delta_2)^2\right]\gamma_1' n_1^{th} + \Gamma^2 \gamma_2' n_2^{th}}{\left[\gamma_1'\gamma_2'/4 - \Gamma^2 - (\omega+\delta_2)(\omega+\delta_1)\right]^2 + \left[(\omega+\delta_1)\gamma_2'/2 + (\omega+\delta_2)\gamma_1'/2\right]^2}, \quad \text{(S12a)}$$

$$S_{n_2}(\omega) = \frac{\left[\gamma_1^{'2}/4 + (\omega+\delta_1)^2\right]\gamma_2' n_2^{th} + \Gamma^2 \gamma_1' n_1^{th}}{\left[\gamma_1'\gamma_2'/4 - \Gamma^2 - (\omega+\delta_2)(\omega+\delta_1)\right]^2 + \left[(\omega+\delta_1)\gamma_2'/2 + (\omega+\delta_2)\gamma_1'/2\right]^2}, \quad \text{(S12b)}$$

where $n_{1,2}^{th}$ is the average effective thermal phonon number and $\delta_{1,2}$ is the effective frequency of two oscillators. Notably, Eqs. (S12a) and (S12b) are only valid for the nonlasing regime. Further, Eqs. (S12a) and (S12b) can be written in a general expression with the superposition of two Lorentz-dispersive lineshapes as

$$S_{n_{1,2}}(\omega) = \frac{A_+(\omega-\omega_+)\gamma_+ + B_+\gamma_+^2}{(\omega-\omega_+)^2 + \gamma_+^2/4} + \frac{A_-(\omega-\omega_+)\gamma_- + B_-\gamma_-^2}{(\omega-\omega_-)^2 + \gamma_-^2/4}. \quad \text{(S13)}$$

Here $\omega_\pm$ and $\gamma_\pm$ are the frequency and linewidth of normal modes of the coupled system. $A_\pm$ and $B_\pm$ are real constants. Therefore, $\omega_\pm$ and $\gamma_\pm$ can be extracted by fitting the measured noise power spectrum.

Figure S3 presents the measured noise power spectra for three distinct regimes and their corresponding data fitting. Figure S3a shows a typical noise power spectrum in the region before the EP, where the real components are distinguishable. Figure S3b is for the region between the EP and lasing threshold, where the real components coalesce and the imaginary components are nondegenerated. We use Eq. (S13) to fit the spectra in these two cases, and the fitting parameters are $\gamma_+/2\pi(\gamma_-/2\pi)$=82.6(75.6) Hz and $\omega_+/2\pi(\omega_-/2\pi)=-13.0(-976.4)$ Hz for Fig. S3a, and $\gamma_+/2\pi(\gamma_-/2\pi)$=40.0(130.0) Hz and $\omega_+/2\pi(\omega_-/2\pi)=-207.4(-210.6)$ Hz for Fig. S3b. Figure

S3c is for the region where the phonon lasing occurs. In this region, we use a single Lorentz function to fit the data, which has $\gamma/2\pi = 3.5$ Hz and $\omega/2\pi = -58.1$ Hz.

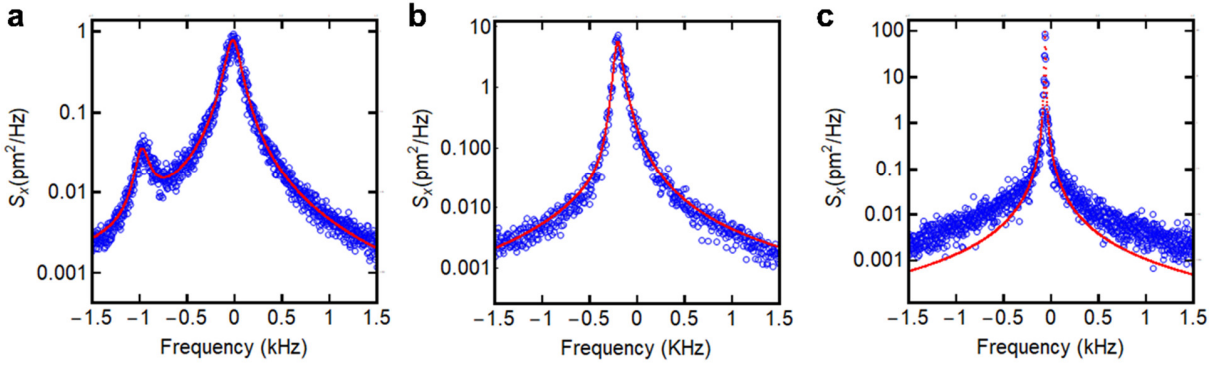

FIG. S3. Data fitting of the measured noise power spectra with coupling strength $\Gamma/2\pi=164$ Hz. The blue circles are the experiment data and the red curves are the corresponding fitting. (a) shows the case before the EP ( $\delta_d/2\pi \approx -1$ kHz ). (b) is between the EP and lasing threshold ( $\delta_d/2\pi \approx -300$ Hz ). (c) is for the phonon lasing regime ( $\delta_d/2\pi \approx 0$ Hz ).

Next, we present how to calibrate phonon numbers and obtain the second-order correlation function. According to the equipartition theorem, i.e., $k_B T_i = m\omega_i^2 \langle x_i^2 \rangle$, the thermal noise spectrum of the fundamental mode at room temperature is used to calibrate the effective phonon number of each mechanical mode. With the real-time measurements of the lock-in amplifier, the mechanical displacement $x_i$ of $i^{th}$ membrane can be reconstructed by the quadrature components $X_i$ and $Y_i$. Thus, the real-time phonon number of $i^{th}$ membrane can be obtained as

$$N_i(t) = A_i \left( X_i(t)^2 + Y_i(t)^2 \right)/2. \quad (i=1,2) \tag{S14}$$

Here $A_i = 2k_B T_{room} / \hbar \omega_i \langle X_i^2 + Y_i^2 \rangle_{room}$ is the calibrated coefficient with $\langle X_i^2 + Y_i^2 \rangle_{room}/2$ being the statistical average of thermal noise signal of fundamental mode at room temperature $T_{room}$. Further, the second-order correlation function of $i^{th}$ mechanical mode in the classical limit can be shown as

$$g_i^{(2)}(\tau) = \frac{\langle N_i(t) N_i(t+\tau) \rangle}{\langle N_i(t) \rangle \langle N_i(t+\tau) \rangle}. \tag{S15}$$

**Supplementary References**

[1] Markus Aspelmeyer, Tobias J. Kippenberg, and Florian Marquardt, Cavity optomechanics, Rev. Mod. Phys. **86**, 1391 (2014).

[2] H. Xu, D. Mason, L. Jiang, and J. G. E. Harris, Topological energy transfer in an optomechanical system with an exceptional point, Nature **537**, 80-83 (2016).

[3] Cheng Yang, Xinrui Wei, Jiteng Sheng, and Haibin Wu, Phonon heat transport in cavity-mediated optomechanical nanoresonators, Nature Communications **11**, 4656 (2020).

[4] H. Seok, L. F. Buchmann, S. Singh, and P. Meystre, Optically mediated nonlinear quantum optomechanics, Phys. Rev. A **86**, 063829 (2012).

[5] L. F. Buchmann, E. M. Wright, and P. Meystre, Phase conjugation in quantum optomechanics, Phys. Rev. A **88**, 041801(R) (2013).